# X-ray diffraction and computer simulation studies of the structure of liquid aliphatic aldehydes: from propanal to nonanal

*Ildikó Pethes\*, László Temleitner, Matija Tomšič, Andrej Jamnik, László Pusztai*

I. Pethes, L. Temleitner, L. Pusztai
Wigner Research Centre for Physics, Hungarian Academy of Sciences,
Budapest, Konkoly Thege út 29-33., H-1121, Hungary

L. Pusztai
International Research Organisation for Advanced Science and Technology (IROAST), Kumamoto University, 2-39-1 Kurokami, Chuo-ku, Kumamoto 860-8555, Japan

M. Tomšič, A. Jamnik
Faculty of Chemistry and Chemical Technology (FKKT), University of Ljubljana,
Večna pot 113, 1000 Ljubljana, Slovenia

Corresponding author: e-mail: pethes.ildiko@wigner.mta.hu

## Abstract

Synchrotron X-ray diffraction experiments and molecular dynamics simulations have been performed on simple aliphatic aldehydes, from propanal to nonanal. The performance of the OPLS all-atom interaction potential model for aldehydes has been assessed via direct comparison of simulated and experimental total scattering structure factors. In general, MD results reproduce the experimental data at least semi-quantitatively. However, a slight mismatch can be observed between the two datasets in terms of the position of the main diffraction maxima. Partial radial distribution functions have also been calculated from the simulation results. Clear differences could be detected between the various O-H partial radial distribution functions, depending on whether the H atom is attached to the carbon atom that is doubly bonded to the oxygen atom of the aldehyde group or not.

## 1. Introduction

Simple aldehydes (a.k.a. alkanals) consist of an alkyl chain and the terminal –CHO group, containing a double covalent bond between C and O as depicted in **Figure 1**. They are rather basic and important organic compounds in the organic chemistry educational curriculum[1] and are exploited in numerous practical and scientific applications (see Ref.[2] and references therein). Some of the simplest aldehydes, as formaldehyde (metanal, H-CHO, one carbon atom), acetaldehyde (etanal, $CH_3$-CHO, two carbon atoms) and butyraldehyde (butanal, $CH_3CH_2CH_2$-CHO, four carbon atoms), are produced by the chemical industry on the scale of million tons/year.[3] Related also to their industrial importance, their thermodynamic properties have been frequently targeted by investigations based on computer simulations in the 21$^{st}$ century.[4,5]

Despite all this, the liquid structure of alkanals has not been investigated too widely: concerning the liquid aldehydes under ambient conditions, there is only one systematic study (at least to the best of our knowledge) that has touched upon the issue of structure.[2] In that work,[2] small- and wide-angle X-ray scattering (SWAXS) data have been measured over a limited scattering vector range, and subsequent computer simulations were performed utilizing the 'united atom' TraPPE force-field model[4]. In this framework, hydrogen atoms in the hydrocarbon molecular tail are not treated as individual sites, but rather as a part of a united-atom site.

In the present work, we extend this limited database on the structure of liquid pure aldehydes, by (1) determining the total scattering structure factor, $F(Q)$, over the widest experimentally available scattering variable range, and (2) by considering also the H-atoms in the hydrocarbon molecular tail as individual single-sites in the subsequent computer simulations. This way, we intend to broaden the pool of research results on the structural observed in similar liquid systems (see, e.g., Ref.[6]).

## 2. Experimental

Synchrotron X-ray diffraction measurements have been carried out at the BL04B2[7] high energy X-ray diffraction beamline of the Japan Synchrotron Radiation Research Institute (SPring-8, Hyogo, Japan). The incoming wavelength was set to 0.2023 Å (corresponding to the energy value of 61.3 keV). This choice has allowed recording the transmission mode diffraction pattern of the samples in the horizontal scattering plane by a single HPGe detector between scattering variable, $Q$, values of 0.16 and 16 Å$^{-1}$ (this is to be compared with the $Q_{max}$ of less than 3 Å$^{-1}$ that could be achieved in Ref.[2]). Samples were placed in 2 mm diameter, thin-walled quartz capillaries (GLAS Müller, Germany); these were mounted in the automatic sample changer. To prevent evaporation of the sample during the measurement, the capillaries were sealed with a cap created from folded Parafilm(R). In order to exploit the dynamic range of the detector optimally, diffraction patterns were recorded in three overlapping frames that differed by the width of incoming beam.

Measured raw intensities were normalized by the incoming beam monitor counts, corrected for absorption, polarization and contributions from the empty capillary. Finally, the patterns over the entire Q-range were obtained by normalizing and merging each frame in electron units, then removing inelastic (Compton) scattering contributions following a standard procedure.[8]

## 3. Molecular dynamics simulations

Classical molecular dynamics (MD) simulations were performed by the GROMACS software package (version 2016.3).[9] The simulations were carried out in cubic simulation boxes with periodic boundary conditions. Simulation boxes contained 2000 molecules that were initially placed into them randomly. The box side lengths were calculated according to the experimental densities and are gathered in **Table 1**.

The simulations were carried out using the OPLS-AA ('Optimized Potentials for Liquid Simulations – All Atom') force field.[10] In this model all atoms, including hydrogens, are treated explicitly; this is another difference from the previous structure study on alkanals.[2] Interactions between atoms are described as the sum of two contributions: the non-bonded and bonded interactions. The former consists of the pairwise additive Coulomb potential, accounting for electrostatics, and the 12-6 Lennard-Jones (LJ) potential for the van der Waals interactions (Equation (1)):

$$V_{ij}^{NB}(r_{ij}) = \frac{1}{4\pi\varepsilon_0} \frac{q_i q_j}{r_{ij}} + 4\varepsilon_{ij}\left[\left(\frac{\sigma_{ij}}{r_{ij}}\right)^{12} - \left(\frac{\sigma_{ij}}{r_{ij}}\right)^6\right], \quad (1)$$

where $r_{ij}$ is the distance between particles $i$ and $j$, $q_i$ and $q_j$ are the partial charges on these particles, $\varepsilon_0$ is the vacuum permittivity, and $\varepsilon_{ij}$ and $\sigma_{ij}$ represent the energy and distance parameters of the LJ potential. The $\varepsilon_{ij}$ and $\sigma_{ij}$ parameters between unlike atoms are calculated according to the geometric combination rules: both parameters are calculated as the geometric average of the homoatomic parameters ($\varepsilon_{ij}=(\varepsilon_{ii}\varepsilon_{jj})^{1/2}$ and $\sigma_{ij}=(\sigma_{ii}\sigma_{jj})^{1/2}$).

Interactions between neighboring atoms within a molecule, i.e. atoms that are covalently bonded (first neighbors), or linked by one (second neighbors) or two atoms (third neighbors) are described by the bonded interactions. First and second neighbors are completely excluded from the non-bonded interactions, while for the third neighbors a reduced LJ and Coulomb potential is applied (reduction is by a factor of 2).

The intramolecular (bonded) forces considered here are the bond stretching (2-body), angle bending (3-body) and the dihedral angle torsion (4-body) interactions. Bond stretching between two covalently bonded $i$ and $j$ atoms is represented by a harmonic potential (Equation (2)):

$$V^b(r_{ij}) = \frac{1}{2} k_{ij}^b (r_{ij} - b_{ij})^2, \quad (2)$$

where $k^b_{ij}$ is the force constant and $b_{ij}$ is the equilibrium distance of the bonded pair. The bond angle vibration of a triplet of atoms $i$, $j$ and $k$ (where atom $j$ is in the middle), is also represented by a harmonic potential (Equation (3)):

$$V^a(\theta_{ijk}) = \tfrac{1}{2} k^a_{ijk}(\theta_{ijk} - \theta^0_{ijk})^2 \tag{3}$$

where $k^a_{ijk}$ is the force constant and $\theta^0_{ijk}$ is the equilibrium angle. The dihedral angle torsion in the OPLS-AA force field is given as the first three terms of a Fourier series (Equation (4)):

$$V_F(\varphi_{ijkl}) = \tfrac{1}{2}\left[F_1\left(1 + \cos(\varphi_{ijkl})\right) + F_2\left(1 - \cos(2\varphi_{ijkl})\right) + F_3\left(1 + \cos(3\varphi_{ijkl})\right)\right], \tag{4}$$

where $F_1$, $F_2$, $F_3$ are force constants, $\varphi_{ijkl}$ is the angle between $ijk$ and $jkl$ planes and $\varphi_{ijkl} = 0$ corresponds to the 'cis' conformation ($i$ and $l$ are on the same side). Actual values of the various parameters are collected in **Table 2, 3 and 4**. For carbon atoms, three different parameter sets are used for the three kinds of carbons in CHO, $CH_2$ and $CH_3$ groups; these carbon atoms are denoted here as C', C and C*, respectively. Two different kinds of hydrogen atoms are distinguished: H stands for hydrogen atoms that belong to $CH_2$ and $CH_3$ groups, while hydrogen atoms in -CHO groups are marked as H'. These notations are defined explicitly in Figure 1.

Two sets of simulations were carried out, (1) with constraining the bonds to their equilibrium lengths (using the LINCS algorithm[12]) and (2) also with flexible bonds; angles and torsional angles were always flexible. Coulomb interactions were treated by the smoothed particle-mesh Ewald (SPME) method,[13] using a 20 Å cutoff in direct space. The short-range van der Waals interactions were also truncated at 20 Å.

Initially, energy minimization was performed using the steepest-descent algorithm. If this was successful then the leap-frog algorithm was applied for integrating Newton's equations of motion, using a 2 fs time step. The calculations were performed at constant volume and temperature (NVT ensemble), at T = 293.15 K. The temperature was kept constant by the Nose-Hoover thermostat,[14] with $\tau_T$ = 2.0. Following a 2 ns equilibration period, particle configurations were collected in every 40 ps between 2 ns and 4 ns. The 'gmx rdf' programme of the GROMACS software was used to calculate the partial radial distribution functions (PRDF), $g_{ij}(r)$. The partial structure factors ($S_{ij}(Q)$) were obtained by Equation (5):

$$S_{ij}(Q) - 1 = \frac{4\pi\rho_0}{Q} \int_0^\infty r\big(g_{ij}(r) - 1\big)\sin(Qr)\mathrm{d}r, \tag{5}$$

where $Q$ is the amplitude of the scattering vector, and $\rho_0$ is the average number density. The X-ray weighted total scattering structure factor ($F(Q)$) is given by Equation (6):

$$F(Q) = \sum_{i \leq j} w_{ij}(Q) S_{ij}(Q), \tag{6}$$

where the $w_{ij}(Q)$ scattering weights can be obtained by Equation (7):

$$w_{ij}(Q) = (2 - \delta_{ij}) \frac{c_i c_j f_i(Q) f_j(Q)}{\sum_{ij} c_i c_j f_i(Q) f_j(Q)}. \tag{7}$$

Here $\delta_{ij}$ is the Kronecker delta, $c_i$ denotes the atomic concentration and $f_i(Q)$ is the atomic form factor, obtainable from parametrized polynomials provided in Reference.[15]

## 4. Results and discussion

Total scattering structure factors from X-ray diffraction experiments are shown in **Figure 2**, together with results obtained from MD simulations. The simulated curves were essentially identical for calculations with either flexible or constrained bonds; only the latter ones are displayed in Figure 2.

At first sight, agreement between measured and calculated structure factors is excellent. However, one can observe some minor discrepancies, notably, for the liquid containing the smallest molecules in the series (propanal, 3 carbon atoms only). Closer inspection reveals that the position of the principal diffraction peak is in general overestimated by MD, by at least 0.02 (butanal), and in most cases, by 0.04 Å$^{-1}$, as shown in **Table 5** and the insert of Figure 2. Similarly, the absolute values of the intensities of the small-angle peak, i.e. the maximum preceding the principal peak (called sometimes fashionably as 'pre-peaks', or 'first sharp diffraction peaks') are not reproduced as well as the rest of the measured signal. We speculate that this situation may be improved if the (somewhat strange) distribution of partial charges over the aldehyde group would be re-considered (cf. Table 2).

The structural relevance of the discrepancies mentioned above could possibly be checked by running RMC_POT calculations that can simultaneously consider the experimental diffraction data and interatomic potential models into account.[16] Such calculations are being planned for one of our future studies. Currently, we can only claim that agreement between experiment and computer simulation is at least semi-quantitative. We are convinced that this is sufficient to justify the following discussion of some of the resulting partial radial distribution functions.

PRDF-s have been determined from the atomic coordinates resulting from the MD-simulation results. The positions of the first maxima have been gathered in **Table 6**. In terms of peak positions, variations over the various aldehydes are negligible. On the other hand, first peak heights for each PRDF presented here grow systematically from propanal to nonanal. This is not due to some important structural change across the series, but mainly reflects a technical issue, the growing molecular hydrocarbon tail (increasing the number of $CH_2$-groups decreases the individual weights of PRDF-s that concern only one specific atom from two molecules, by decreasing the partial number density of the specific atoms in question.)

The main target at present are the various O-H partials, with the aim to see if either of them deviates from the others, particularly the O-H' one. In **Figure 3**, the three $g_{OH}(r)$ functions are presented. The O-H(C*) curve is distinctly different from the other two, which suggests that the 'aldehyde-ends' of the molecules do not prefer to be in the close vicinity of the C* carbon atom of the surrounding molecules. This notion is supported by the O-Hα and O-H' PRDF-s, the latter showing stronger preference for neighboring than the former. As readable from Table 6, all O-H first neighbor distances are much longer than a usual hydrogen-bonding distance (which would be at most 2 Å, see, e.g., Ref.[17]); this concept therefore is not even distantly applicable for aldehydes.

Pair correlations between some of the heavy atoms are represented by the O-O and C(CHO)-C(CHO) PRDF-s, see **Figure 4**. The first two curves are quite similar to each other, with essentially identical maximum positions. This suggests that no preferred spatial arrangements (orientational correlations) are present between neighboring aldehyde groups. Although the shape of the C*C*

PRDF-s is somewhat different from that of the C'C' ones, the contrast, in terms of peak intensity, is much weaker than it was for between the O-H' and O-H(C*) PRDF-s. That is, the different C-C PRDF-s do not contradict with the conjecture that the 'aldehyde-ends' of the molecules prefer the vicinity of each other.

**5. Summary and Conclusions**

New synchrotron X-ray diffraction experiments have been performed over a wide range of the scattering vector for a series of alkanals (alkyl-aldehydes), from propanal (3 C atoms) to nonanal (9 C atoms). To the best of our knowledge, this is the first such a systematic set of wide angle diffraction data for simple aldehydes. Therefore, data presented here may serve as future reference on the structure of liquid alkanals.

Molecular dynamics computer simulations with all-atom force-field models (including also hydrogen atoms as distinct sites) have been carried out, as complementary to the diffraction measurements. At least semi-quantitative agreement with measured total scattering structure factors is observed. There are slight discrepancies observed only in the position of the main maximum, which could possibly be minimized by re-considering partial charges in the aldehyde group.

Based on the 3 different O-H PRDF-s, it may be suggested that neighboring molecules turn toward each other (somewhat) preferentially by their aldehyde ends. From $g_{OO}(r)$ and $g_{C'C'}(r)$ it may be discerned that no (or at most, very weak) specific orientational correlations are present between neighboring aldehyde groups.

As far as possible improvements concerning agreement with experimental total scattering structure factors is concerned, we are planning to perform RMC_POT calculations, in order to establish whether the discrepancies reported in the previous section are important and would change any of the structural findings presented here.

**Acknowledgment**

The authors are grateful to the National Research, Development and Innovation Office (NKFIH) of Hungary for financial support through Grants No. SNN 116198, TÉT_16-1-2016-0056, and to the Slovenian Research Agency for the financial support through the research core funding No. P1-0201 and the project No. N1-0042 (Structure and thermodynamics of hydrogen-bonded liquids: from pure water to alcohol-water mixtures).

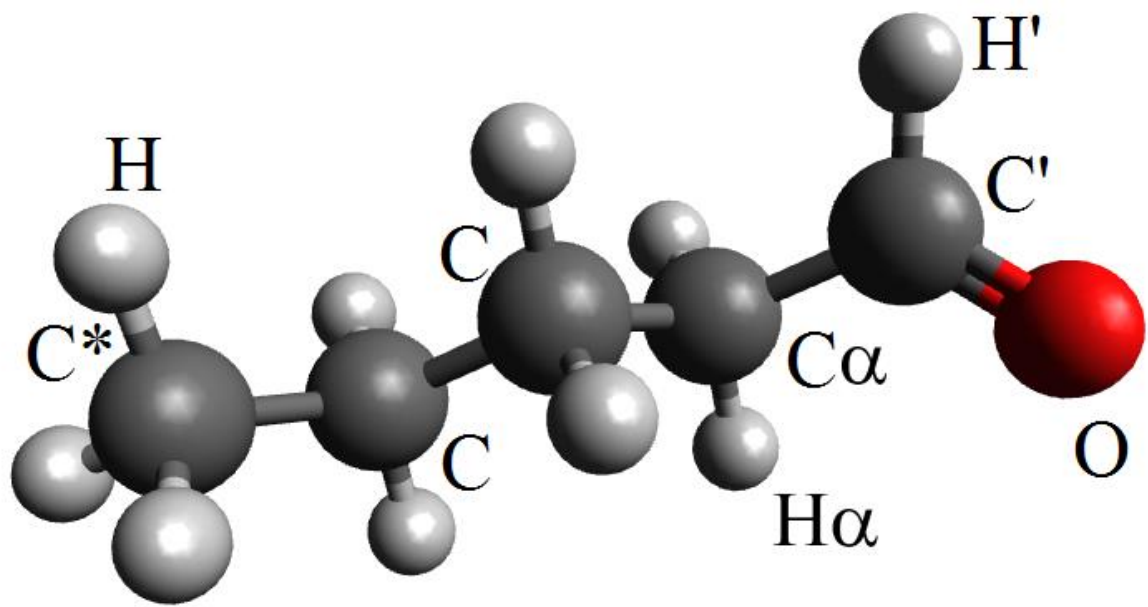

**Figure 1.** Schematic representation of a pentanal molecule in the all-atom model. Carbon atoms in -CHO, -CH$_2$- and -CH$_3$ groups are denoted as C', C and C*, respectively. Hydrogen atoms that belong to -CH$_2$- and -CH$_3$ groups are marked as H, while hydrogen atoms in -CHO groups are denoted as H'. The carbon and hydrogen atoms in the -CH$_2$- group nearest to the -CHO group are marked as Cα and Hα, respectively.

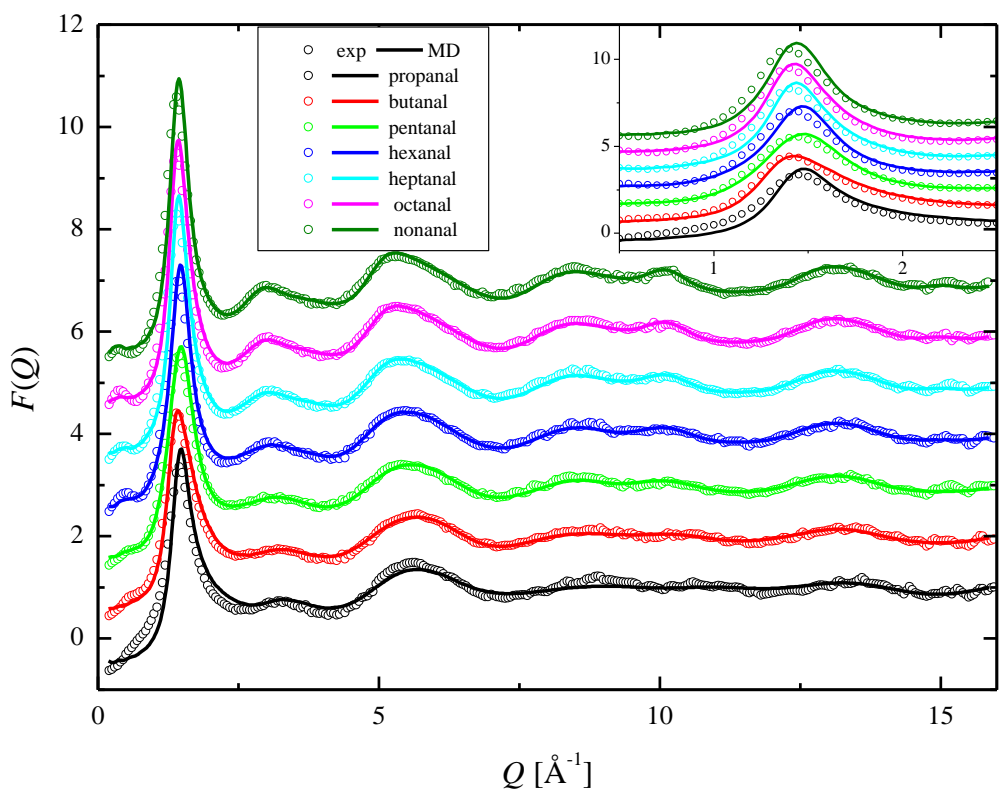

**Figure 2.** Experimental (symbols) and simulated (lines) X-ray diffraction structure factors for propanal (black), butanal (red), pentanal (light green), hexanal (blue), heptanal (cyan), octanal (magenta) and nonanal (dark green). For the sake of clarity, the curves from propanal to nonanal are shifted upwards by a constant factor of +1. Inset: zoom to the main maximum.

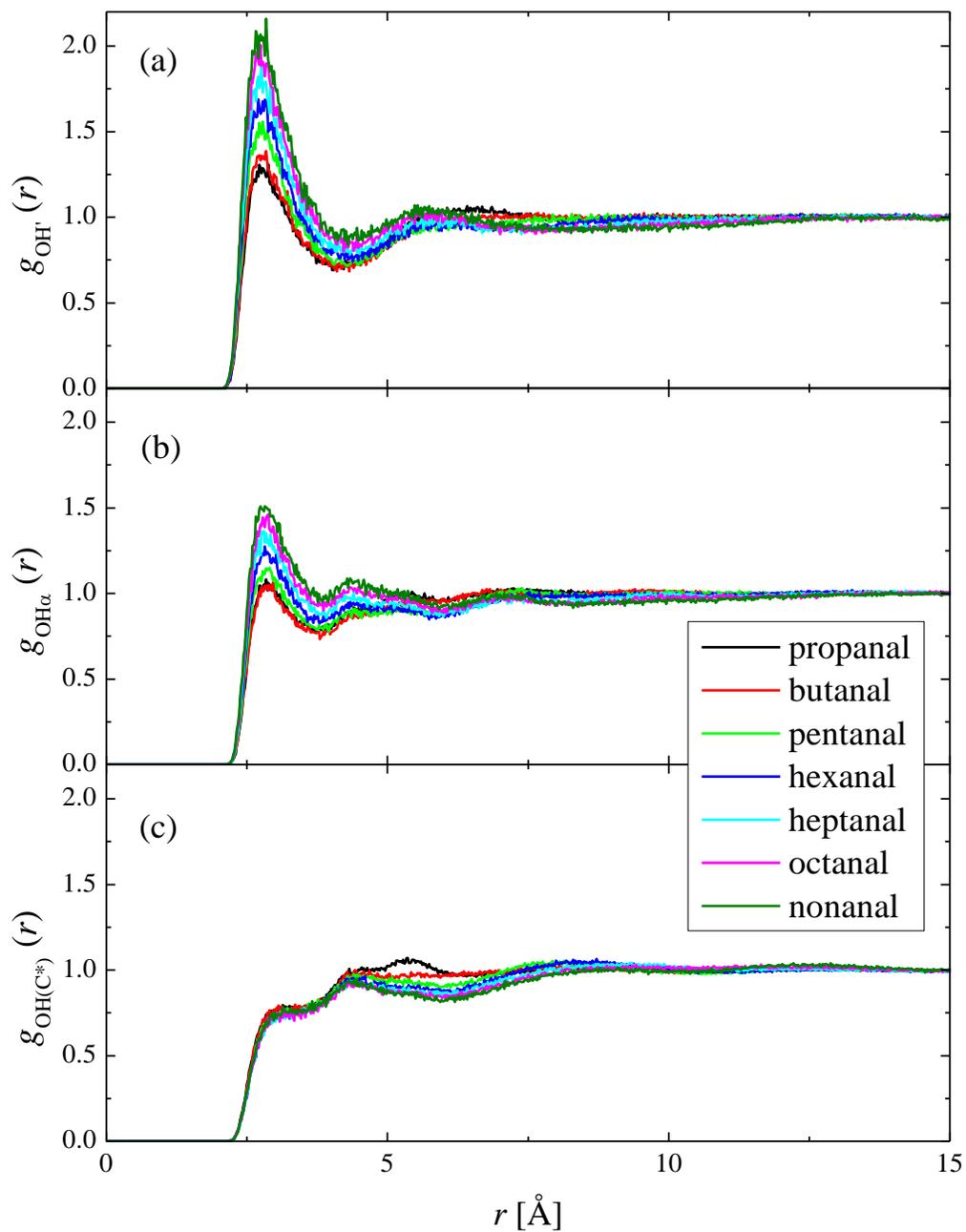

**Figure 3.** (a) O-H', (b) O-Hα and (c) O-H intermolecular partial radial distribution functions obtained from MD simulations for propanal (black), butanal (red), pentanal (light green), hexanal (blue), heptanal (cyan), octanal (magenta) and nonanal (dark green).

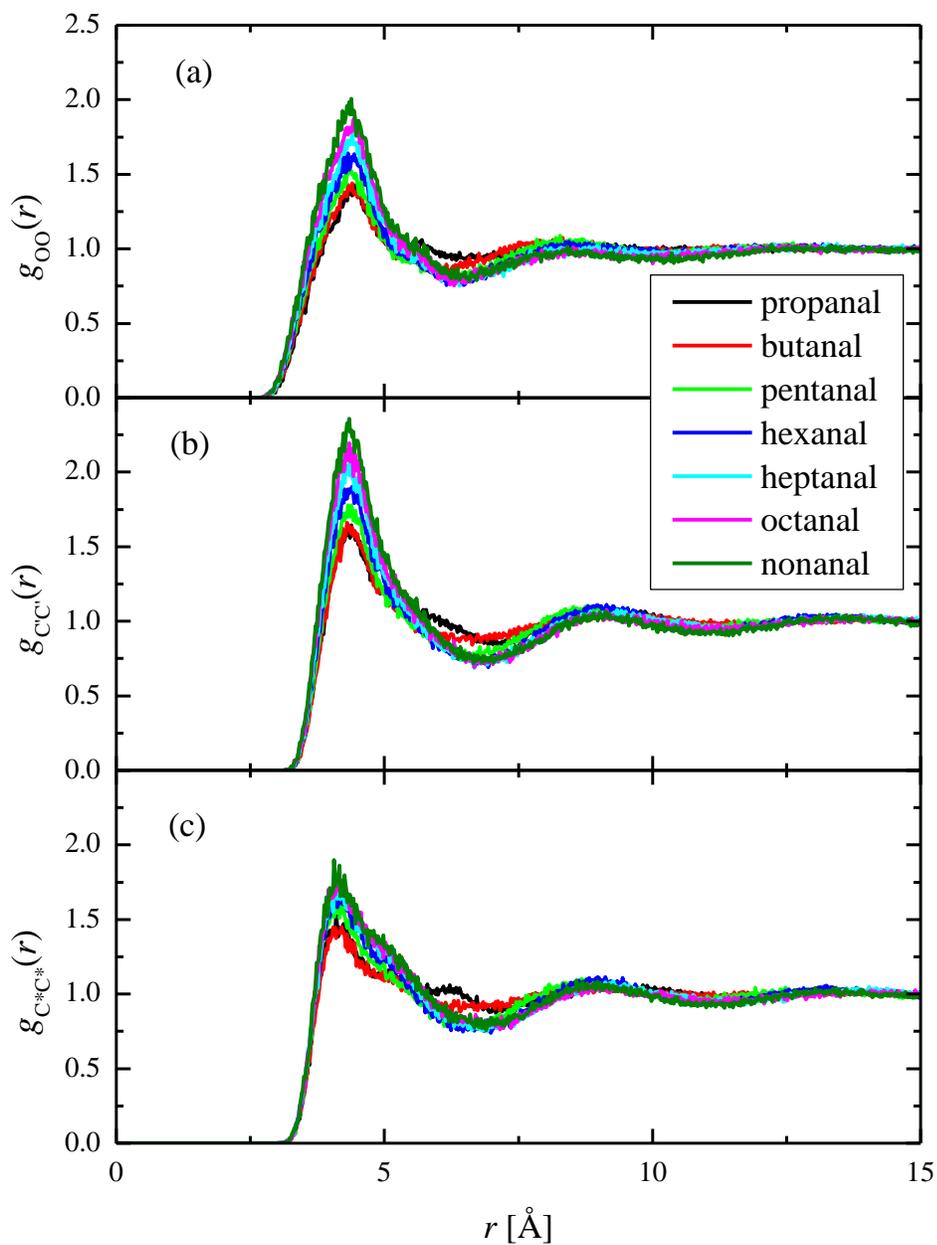

**Figure 4.** (a) O-O, (b) C'-C' and (c) C*-C* partial radial distribution functions obtained from MD simulations for propanal (black), butanal (red), pentanal (light green), hexanal (blue), heptanal (cyan), oxanal (magenta) and nonanal (dark green).

**Table 1.** Densities, number densities and the sizes of the simulation boxes for the aldehydes investigated. The simulation boxes contained 2000 molecules.

|  | Molecular formula | Density [g cm$^{-3}$] | Number density [Å$^{-3}$] | Box length [nm] |
|---|---|---|---|---|
| Propanal | $CH_3CH_2CHO$ | 0.805 | 0.083467 | 6.21115 |
| Butanal | $CH_3(CH_2)_2CHO$ | 0.8 | 0.086856 | 6.68946 |
| Pentanal | $CH_3(CH_2)_3CHO$ | 0.81 | 0.09061 | 7.06847 |
| Hexanal | $CH_3(CH_2)_4CHO$ | 0.815 | 0.093101 | 7.41781 |
| Heptanal | $CH_3(CH_2)_5CHO$ | 0.817 | 0.094791 | 7.74274 |
| Octanal | $CH_3(CH_2)_6CHO$ | 0.82 | 0.096285 | 8.03780 |
| Nonanal | $CH_3(CH_2)_7CHO$ | 0.827 | 0.098035 | 8.29729 |

**Table 2.** Non-bonded parameters (partial charges and LJ parameters) of the aldehydes used in MD simulations.[10] Notations of the atoms are defined in Figure 1.

| Atom | $q$ [e] | $\sigma_{ii}$ [nm] | $\varepsilon_{ii}$ [kJ mol$^{-1}$] |
|---|---|---|---|
| C' | 0.45 | 0.375 | 0.43932 |
| C | -0.12 | 0.35 | 0.276144 |
| C* | -0.18 | 0.35 | 0.276144 |
| O | -0.45 | 0.296 | 0.87864 |
| H' | 0.0 | 0.242 | 0.06276 |
| H | 0.06 | 0.25 | 0.12552 |

**Table 3.** Bond stretching and angle bending parameters used in the MD simulations.[10]

| $ij$ or $ijk$ | $b_{ij}$ [nm] or $\theta^0_{ijk}$ [degree] | $k^b_{ij}$ [kJ/(mol nm$^2$)] or $k^a_{ijk}$ [kJ mol$^{-1}$ rad$^{-2}$)] |
|---|---|---|
| H(H')-C(C*,C') | 0.109 | 284512 |
| O-C' | 0.1229 | 476976 |
| C(C*, C')-C | 0.1529 | 224262.4 |
| H-C(C*)-H | 107.8 | 276.144 |
| H-C(C*)-C(C*) | 110.7 | 313.8 |
| C(C*)-C-C | 112.7 | 488.273 |
| H'-C'-C | 115.0 | 292.88 |
| H'-C'-O | 123.0 | 292.88 |
| C-C'-O | 120.4 | 669.44 |
| C(C*)-C-C' | 111.1 | 527.184 |
| H-C-C' | 109.5 | 292.88 |

**Table 4.** Dihedral angle torsion force constants used in MD simulations.[10]

| ijkl | $F_1$ [kJ mol$^{-1}$] | $F_2$ [kJ mol$^{-1}$] | $F_3$ [kJ mol$^{-1}$] |
|---|---|---|---|
| H-C(C*)-C-H [a] | 0 | 0 | 1.2552 |
| H-C(C*)-C-C(C*) [a] | 0 | 0 | 1.2552 |
| C(C*)-C-C-C [a] | 5.4392 | -0.2092 | 0.8368 |
| H'-C'-C-C(C*) | 0 | 0 | 0 |
| H'-C'-C-H | 0 | 0 | 1.50624 |
| O-C'-C-C(C*) | -1.159 | 5.137952 | -2.903696 |
| O-C'-C-H | 0 | 0 | 0 |
| C'-C-C-C(C*) | -7.1002 | -1.907904 | 2.44764 |
| C'-C-C(C*)-H | 0 | 0 | -0.317984 |

[a] These parameters of the original OPLS-AA force field[10] were modified according to Ref. [11].

**Table 5.** Positions of the first peak of the $F(Q)$ curves obtained from experiment and MD simulation (in Å$^{-1}$).

| aldehyde | Experiment | Simulation |
|---|---|---|
| propanal | 1.45 | 1.48 |
| butanal | 1.40 | 1.42 |
| pentanal | 1.44 | 1.48 |
| hexanal | 1.43 | 1.47 |
| heptanal | 1.40 | 1.44 |
| octanal | 1.39 | 1.43 |
| nonanal | 1.40 | 1.44 |

**Table 6.** Positions of the first intermolecular peak of selected $g_{ij}(r)$ curves obtained in MD simulations (in Å).

| Pair ($ij$) | $r$ [Å] |
| --- | --- |
| O-H' | 2.8 |
| O-H$\alpha$ | 2.8 |
| O-H (H in CH$_3$) | ca. 3.0 |
| O-O | 4.4 |
| C'-C' | 4.3 |